# $KH_2PO_4$ + Host Matrix (Alumina / $SiO_2$) Nanocomposite: Raman Scattering Insight

Ya. Shchur[1], A.S. Andrushchak[2], V.V. Strelchuk[3], A.S. Nikolenko[3], V.T. Adamiv[4], N.A. Andrushchak[2], P. Göring[5], P. Huber[6], A.V. Kityk[7]

[1]*Institute for Condensed Matter Physics, 1 Svientsitskii str., 79011, Lviv, Ukraine*
[2]*Lviv Polytechnic National University, 12 S. Bandery str., 79013, Lviv, Ukraine*
[3]*V.E. Lashkaryov Institute of Semiconductor Physics, National Academy of Sciences of Ukraine, 41 Prosp. Nauky, 03028 Kyiv, Ukraine*
[4]*O.G. Vlokh Institute of Physical Optics, 23 Dragomanova str., 79005 Lviv, Ukraine*
[5]*SmartMembranes GmbH, Heinrich-Damerow-Str.4, 06120 Halle, Germany*
[6]*Institute of Materials Physics and Technology, Hamburg University of Technology, Eissendorferstr. 42, D-21073 Hamburg, Germany*
[7]*Faculty of the Electrical Engineering, Czestochowa University of Technology, Al. Armii Krajowej 17, 42-200, Czestochowa, Poland*
*Tel: (032) 2761978, Fax: (032) 2761158, e-mail: shchur@icmp.lviv.ua*

**ABSTRACT**
We report on the synthesis and Raman scattering characterization of composite materials based on the hostnanoporous matrices filled with nanostructured $KH_2PO_4$ (KDP) crystal. Silica ($SiO_2$) and anodized aluminium oxide (AAO) were used as host matrices with various pore diameters, inter-pore spacing and morphology. The structure of the nanocomposites was investigated by X-ray diffraction and scanning electron microscopy. Raman scattering reveals the creation of one-dimensional nanostructured KDP inside the $SiO_2$ matrix. We clearly observed the stretching $\nu_1$, $\nu_3$ and bending $\nu_2$ vibrations of $PO_4$ tetrahedral groups in the Raman spectrum of $SiO_2$ + KDP. In Raman scattering spectra of AAO + KDP nanocomposite, the broad fluorescence background of AAO matrix dominates to a great extent, hindering thus the detecting of the KDP compound spectral response.

## 1. INTRODUCTION

For many decades, KDP crystal has been a significant technological compound intensively utilized in quantum electronic devices, e.g. as second- and third-optical harmonic generator, laser Q-switcher, electro-optical deflector etc. However, in all these applications KDP was used as a bulk three-dimensional (3D) single crystal. There appears the question of how may evolve the main physical properties of KDP going from 3D to one-dimensional (1D) space morphology? To elucidate this issue we created nanocomposite materials consisting of nanoporous host matrices and 1D nanorods of KDP crystals synthesized inside the pores.

It is worth noting that the second-harmonic generation was already detected in KDP + AAO nanocomposite [1, 2]. Clear manifestation of nanoscale confinement effect in the optical properties of hydrogen bonded $PbHPO_4$ crystal of KDP type was recently shown in paper [3].

## 2. EXPERIMENTAL

We applied the method of crystal growth in nanoporous matrix by imbibing the pores with saturated solutions of KDP (KOH and $H_3PO_4$). Nanoporous anodized AAO with matrix thickness of $h$ ~205 μm, inter-pore spacing $l$ ~125 nm and pore diameter $d$ ~70-80 nm and Si matrices ($h$ ~200 μm, $d$ ~1 μm) commercially available at SmartMembrans GmbH (Halle, Germany) and nanoporous $SiO_2$ ($h$ ~200 μm, $d$ ~10 nm) material were utilized as the host matrices. The host matrices were normally soaked in saturated solution of KDP at 323 K for 1 hour and subsequently dried at room temperature for a few hours. The pore diameters of the nanocomposite material became smaller compared with those of the pure host matrix. One may suggest the following eventual route of the creation of KDP nanostructured crystals in nanoporous matrix. At the initial stage of matrix soaking into the saturated KDP solution, some KDP seeds appear close to the pore walls forming thus KDP nanotubes with hollow inner diameter. Then the cylindrical-shape nanotubes grow perpendicular to the sample surface. Nanotubes are isolated from each other and do not create the common aggregation on the sample surface.

The sample quality was tested by X-ray diffraction and scanning electron microscopy (SEM). The sample surfaces were mechanically cleaned to eliminate the KDP microcrystals synthesized on the sample surfaces. Only the samples with flat and comparatively uniform surfaces were selected for spectroscopic measurements.

Micro-Raman investigations were performed at room temperature in backscattering configuration using a 785 nm laser as a light source and micro-Raman microscope based on Princeton Instruments IsoPlane SCT-320 spectrograph and liquid nitrogen cooled SI CCD detector (PyLoN 400BR eXcelon, Princeton Instruments). The exciting radiation was focused on the sample surface with ×50/NA 0.75 optical objective, giving a laser spot diameter of ~1 μm. The laser power on the sample surface was varied in the range of 1−2 mW. Raman spectra were registered at different points of the sample (referred as p1, p2 etc) and from both sides of the sample (referred as side1 and side2).

## 3. RESULTS and DISCUSSION

At room temperature, KDP crystallizes in tetragonal symmetry (I-42d, #122). In general features, the Raman spectrum of bulk KDP is typical for the whole class of hydrogen-bonded crystals of KDP-type, i.e. $RbH_2PO_4$, $CsH_2PO_4$, $TlH_2PO_4$, $PbHPO_4$ and their deuterated analogues. Raman spectra of this type crystals can be separated into four distinct regions [4, 5]. The low-frequency range below ~250 cm$^{-1}$ corresponds to the external lattice vibrations of K and $PO_4$ (or $H_2PO_4$) groups, intermediate-frequency region within ~320-600 cm$^{-1}$ contains mostly the internal $\nu_2$, $\nu_4$ bending vibrations of tetrahedral $PO_4$ groups, the high-frequency range of ~900-1050 cm$^{-1}$ comprises the strong covalently bonded internal $\nu_1$ and $\nu_3$ stretching vibrations of $PO_4$ groups and finally all the high frequency modes ($\nu$ > 1200 cm$^{-1}$) should correspond to hydrogen bond O – H…O vibrations, high order Raman scattering and Fermi resonance effects in the spectra. However, according to lattice dynamics simulation [4-6] hydrogen vibrations are often mixed with internal $PO_4$ stretching or even bending vibrations. Note that in aqueous solution the internal modes of the free $PO_4$ tetrahedra reveal the following frequencies, $\nu_2$ = 420 cm$^{-1}$, $\nu_4$ = 560 cm$^{-1}$, $\nu_1$ = 940 cm$^{-1}$, $\nu_3$ = 1020 cm$^{-1}$ [7]. In some peculiar crystal environment, these frequencies may be changed by internal crystal fields. Especially this concerns the low-frequency bending $\nu_2$, $\nu_4$ modes. They may interact with lattice or hydrogen bond vibrations.

Figure 1 presents the Raman spectra recorded at room temperature on the pure host nanoporous $SiO_2$ matrix and nanocomposite material $SiO_2$ + KDP. Despite the clearly distinguishable peaks near ~500, 600, 800, 1000 cm$^{-1}$, the Raman spectrum of silica matrix exhibits a very strong, broad band centered near 400 cm$^{-1}$. All other Raman scattering lines detected in the spectra of nanocomposite $SiO_2$ + KDP compound should be treated as those inherent for the newly synthesized material. As seen in Fig.1, all three spectra taken at different places of nanocomposite sample show a good reproducibility of the main spectral features. Very strong peaks at 904 and 961 cm$^{-1}$ correspond to full-symmetric $\nu_1$ covalently bonded stretching vibrations of the $PO_4$ groups. This $\nu_1$ mode is usually observable in the Raman spectrum of bulk KDP near 900-915 cm$^{-1}$ as a single line. However, due to the nanoscale confinement effect, crystal non-stoichiometry, elastic strains occurred in nanotubes *etc* the frequency degeneracy of $\nu_1$ may be lifted and this mode is probably spitted into two modes, at 904 and 961 cm$^{-1}$ [8]. The same arguments are valid for the medium intensity 1070 and 1174 cm$^{-1}$ modes which are most probably related to the internal $\nu_3$ vibration. This $\nu_3$ mode is normally detected near 1000-1100 cm$^{-1}$ in the bulk KDP as a shoulder of the strongest $\nu_1$ mode at ~900-915 cm$^{-1}$. In our experiment, we observed a broad shoulder centered near 1070 cm$^{-1}$ typical for the bulk KDP and another clearly distinguishable 1174 cm$^{-1}$ line which is evidently the result of nanocrystal confinement effect or, alternatively, the mixing of internal PO$_4$ and hydrogen bond vibrations. Note that according to *ab initio* calculation and *eigen*-vector analysis performed for the bulk KDP crystal the modes placed within the 1031-1128 cm$^{-1}$ range are exactly the result of such mode mixing [6].

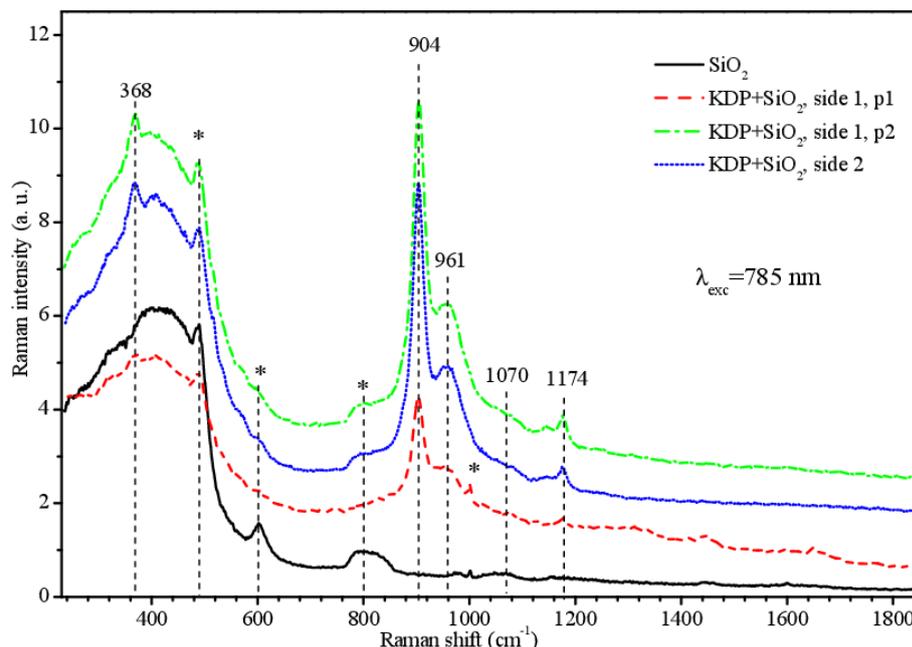

Figure 1. Raman spectra of the host nanoporous $SiO_2$ matrix and nanocomposite material $SiO_2$ + KDP taken from both sides and different points of the sample. Phonon modes inherent for silica spectrum are denoted by asterisks.

In the bulk KDP, PO$_4$ internal bending $\nu_2$ and $\nu_4$ vibrations are localized in ~350-580 cm$^{-1}$ range. Unfortunately, in our nanocomposite SiO$_2$ + KDP sample, this range is overlapped by the strong broad band of the host silica matrix largely hindering the observation of the most of $\nu_2$ and $\nu_4$ vibrations of KDP. However, one strong mode near 368 cm$^{-1}$ is visible on top of the strong silica background. In Raman scattering experiments performed in the past on the bulk KDP, the modes located near 360-370 cm$^{-1}$ were the strongest ones among the set of bending $\nu_2$ and $\nu_4$ vibrations. They were detected either as a singlet at 362 cm$^{-1}$ or as a doublet of two 354 and 386 cm$^{-1}$ modes [8].

We did not succeed in finding the low frequency KDP external lattice modes. Partly this may be explainable by the comparatively low intensity of these modes and partly by the fact that this frequency range is covered by the strong SiO$_2$ scattering background.

The Raman spectra recorded on the host nanoporous AAO matrix and spectra taken from composite KDP + AAO nanostructured material are depicted in Figure 2. As seen from this figure, the Raman spectrum of such composite compound is much poorer compared with those of the SiO$_2$ + KDP material. The main feature of the host AAO matrix is a strong and very broad fluorescence background which significantly hinders the detection of Raman signal from embedded there nanostructured KDP crystals. Only one Raman line at 1015 cm$^{-1}$ typical of the bulk KDP was detected in the spectrum of newly synthesized AAO + KDP compound. This line probably corresponds to the $\nu_3$ stretching PO$_4$ mode. However, why the another stretching $\nu_1$ mode which is usually much stronger in the Raman spectrum of the bulk KDP than the $\nu_3$ mode is not visible in our present spectrum of AAO + KDP is unclear.

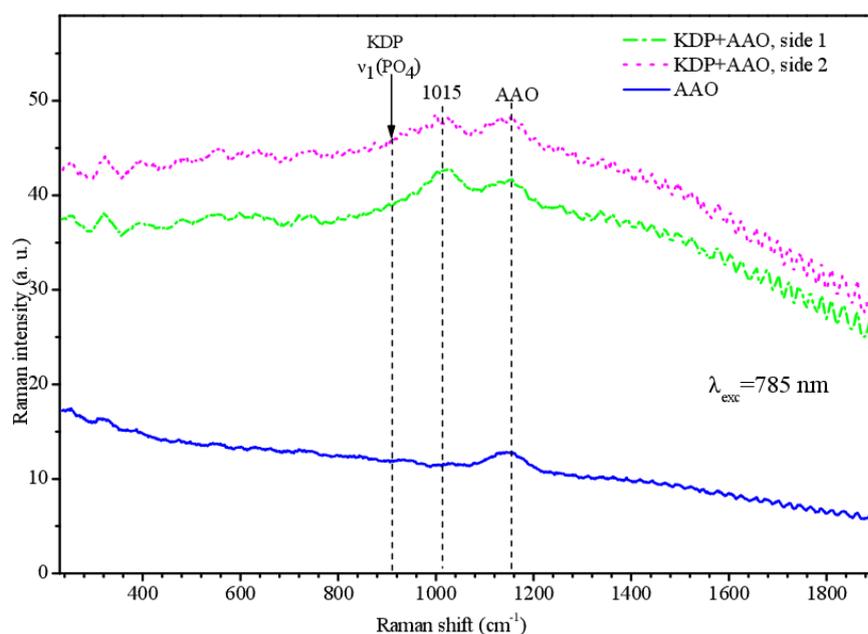

Figure 2. Raman spectra of the host nanoporous AAO matrix and nanocomposite material AAO + KDP taken from both sides of the sample.

## 4. CONCLUSIONS

We synthesized nanocomposite materials consisting of the nanoporous host matrices, SiO$_2$ and AAO filled by nanostructured KDP. Internal vibrations typical of KDP crystals were clearly observed in Raman scattering spectra of SiO$_2$ + KDP nanocomposites corroborating thus the existence of nanosized KDP inside the matrix pores. Due to the broad fluorescence of the host AAO matrix, the Raman spectrum of AAO + KDP nanocomposite is much poorer compared with those of SiO$_2$ + KDP. A single absorption line at 1015 cm$^{-1}$ might be treated the internal vibration of KDP crystal. Further studies are definitely needed to improve the synthesis procedure of nanocomposite materials to get stable, reproducible and high quality nanostructured compounds.

**ACKNOWLEDGEMENTS**

The presented results are part of a project that has received funding from the European Union's Horizon 2020 research and innovation programme under the Marie Skłodowska-Curie grant agreement No 778156. The support from Ministry of Education and Science of Ukraine (project 0119U002255 "Nanocrystallite") is acknowledged. A.V. Kityk acknowledges a support from resources for science in years 2018-2022 granted for the realization of international co-financed project Nr W13/H2020/2018 (Dec. MNiSW 3871/H2020/2018/2). Ya. Shchur was supported by project 0118U003010 of National Academy of Sciences of Ukraine. P. Huber profited from funding by the Deutsche Forschungsgemeinschaft (DFG, German Research Foundation) – Projektnummer 192346071 – SFB 986.